\documentstyle[aps,rotate,multicol,epsf]{revtex}

\begin{document}
\draft

\title{Tracer Dispersion in Porous Media with Spatial Correlations}

\author{Hern\'an A. Makse $^{1-3}$, Jos\'e S. Andrade Jr. $^1$, 
and H. Eugene Stanley $^3$}
\address{$^1$Departamento de F\'{\i}sica, Universidade Federal do Cear\'a,
60451-970 Fortaleza, Cear\'a, Brazil \\
$^2$ Schlumberger-Doll Research, Old Quarry Road, Ridgefield, CT 06877 \\
$^3$ Center for Polymer Studies and Physics Dept., Boston University,
Boston, MA 02215}

\date{\today}
\maketitle
\begin{abstract}
We analyze the transport properties of a neutral tracer in a carrier
fluid flowing through percolation-like porous media with spatial correlations.
We model convection in the mass transport process using the 
velocity field obtained by the numerical solution of the Navier-Stokes and 
continuity equations in the pore space. 
We show that the
resulting statistical properties of the tracer can be approximated by a
L\'evy walk model, which is a consequence of the broad distribution of
velocities plus the existence of spatial correlations in the porous
medium.
\end{abstract}
%We model convection in the tracer dispersion process using the 
%velocity field obtained by the numerical solution of the Navier-Stokes and 
%continuity equations in the pore space. Due to the existence of stagnant 
%zones,
%the tracer is trapped in regions of small fluid velocity. This trapping is 
%statistically relevant when convection is the dominant mechanism of mass 
%transport. We show that, under these conditions, the resulting statistical 
%properties of the tracer can be approximated by the L\'evy statistics, as 
%opposed to the standard Gaussian behavior.

%\pacs{PACS}

\begin{multicols}{2}
\narrowtext
The phenomenon of hydrodynamic dispersion---the unsteady transport of a
neutral tracer in a carrier fluid flowing through a porous material
---has been widely investigated in the fields of petroleum and chemical
engineering ~\cite{saffman,review}. One can identify different regimes
of tracer dispersion according to the P\'eclet number ${\rm Pe}\equiv
v\ell/D_m$, which is the ratio between the typical time for diffusion
$\ell^2/D_m$ and the typical time for convection $\ell /v$.  Here $v$ is
the velocity of the carrier fluid, $\ell$ a characteristic length scale
of the porous media, and $D_m$ the molecular diffusivity of the tracer.

In the small P\'eclet number regime, molecular diffusion dominates the
way in which the tracer samples the flow field.  In the large P\'eclet
number regime, also called {\it mechanical dispersion}, convection
effects are significant; the tracer velocity is approximately equal to
the carrier fluid velocity, and molecular diffusion plays little role.
The tracer samples the disordered medium by following the velocity
streamlines. In a random walk picture, we may think of a tracer particle
following the direction of the velocity field, and taking steps of
length $\ell$ and duration $\ell/v$.

The classical approach to model dispersion in porous media is to
consider microscopically disordered and macroscopic isotropic and
homogeneous porous materials. Under these conditions, dispersion is said
to be {\it Gaussian} and the phenomenon can be mathematically
represented in terms of the convection-diffusion equation \cite{review}.
This traditional formalism, which is valid for Euclidean geometries,
cannot be adopted to describe the global behavior of hydrodynamic
dispersion in heterogeneous systems. Specifically, in the case of
percolation porous media, the breakdown of the macroscopic
convective-diffusion description is a direct consequence of the
self-similar characteristic of the void space geometry.

Here we discuss the interesting physics that arises when the tracer
moves in a flow field with a very broad velocity distribution.
Consider, e.g., fluid flow in percolation clusters near the percolation
threshold---a model system relevant to a porous medium with stagnant
small-velocity zones that are linked with large-velocity zones.  In this
case the typical time for convection $\ell/v$ is without bound since the
velocity can be arbitrarily small in some fluid elements of the void
space. Saffman showed
\cite{saffman} that the mean square duration of a tracer step
is not finite but diverges logarithmically unless an upper cut-off is
introduced into the typical time step. This upper cut-off is imposed by
the mass transport mechanism of molecular diffusion.

Molecular diffusion is expected to affect the tracer motion in two ways
\cite{saffman}:

%\begin{itemize}
{\bf (i)} A quantity of material may cross from one streamline with
fluid velocity $v$ to another by {\it lateral} diffusion if the time
step for convection $\ell_{\parallel} /v$ is larger than $t_1$, where
$t_1= \ell_{\perp}^2/2D_m$ is the characteristic time for molecular
diffusivity effects to become appreciable \cite{taylor} and
$\ell_{\parallel}$ and $\ell_{\perp}$ are the longitudinal and lateral
pore lengths, respectively (with respect to the flow direction)
\cite{saffman}. Thus, if $\ell_{\parallel}/ v \gg t_1$, the tracer has
enough time to diffuse across the pore, and the time step associated
with such a move is $\Delta t = t_1$.  When $\ell_{\parallel}/ v \ll
t_1$, the time duration of a convective step is smaller than the time
required for molecular diffusion, and the tracer moves with the carrier
fluid taking a step of duration $\Delta t = \ell_{\parallel}/v$.

{\bf (ii)} An amount of material may be transported by diffusion {\it
along the pore}. The same considerations as in point {\bf (i)} lead to a
time step $\Delta t = \ell_{\parallel}/v$ in which convection dominates
when $\ell_{\parallel}/ v \ll t_0=\ell_{\parallel}^2/2 D_m$.  Here the
typical length scale is the longitudinal length of the pore
$\ell_{\parallel}$. If $\ell_{\parallel}/ v \gg t_0$, diffusion
dominates and the tracer takes a time step $\Delta t = t_0$.
%\end{itemize}

Here we propose a model of tracer dispersion in a porous medium. The
porous medium is composed of blocks of impermeable material that occupy,
with a given probability $p$, a square lattice. We consider a lattice at
the site percolation threshold, so an incipient spanning cluster is
formed that connects the two ends of the lattice. Previous studies
modeled the convective local ``bias'' for the movement of the neutral
tracer in the porous media assuming Stokes flow \cite{review}. Even at
macroscopically small Reynolds conditions, this assumption might be
violated in real flow through porous media, specially in the case of
heterogeneous materials (e.g., percolation-like structures) where a
broad distribution of pore sizes can lead to a broad distribution of
local fluxes.
\begin{figure}
\centerline{
 (\bf{a})
\epsfxsize=4.2cm \rotate[r]{\epsfbox{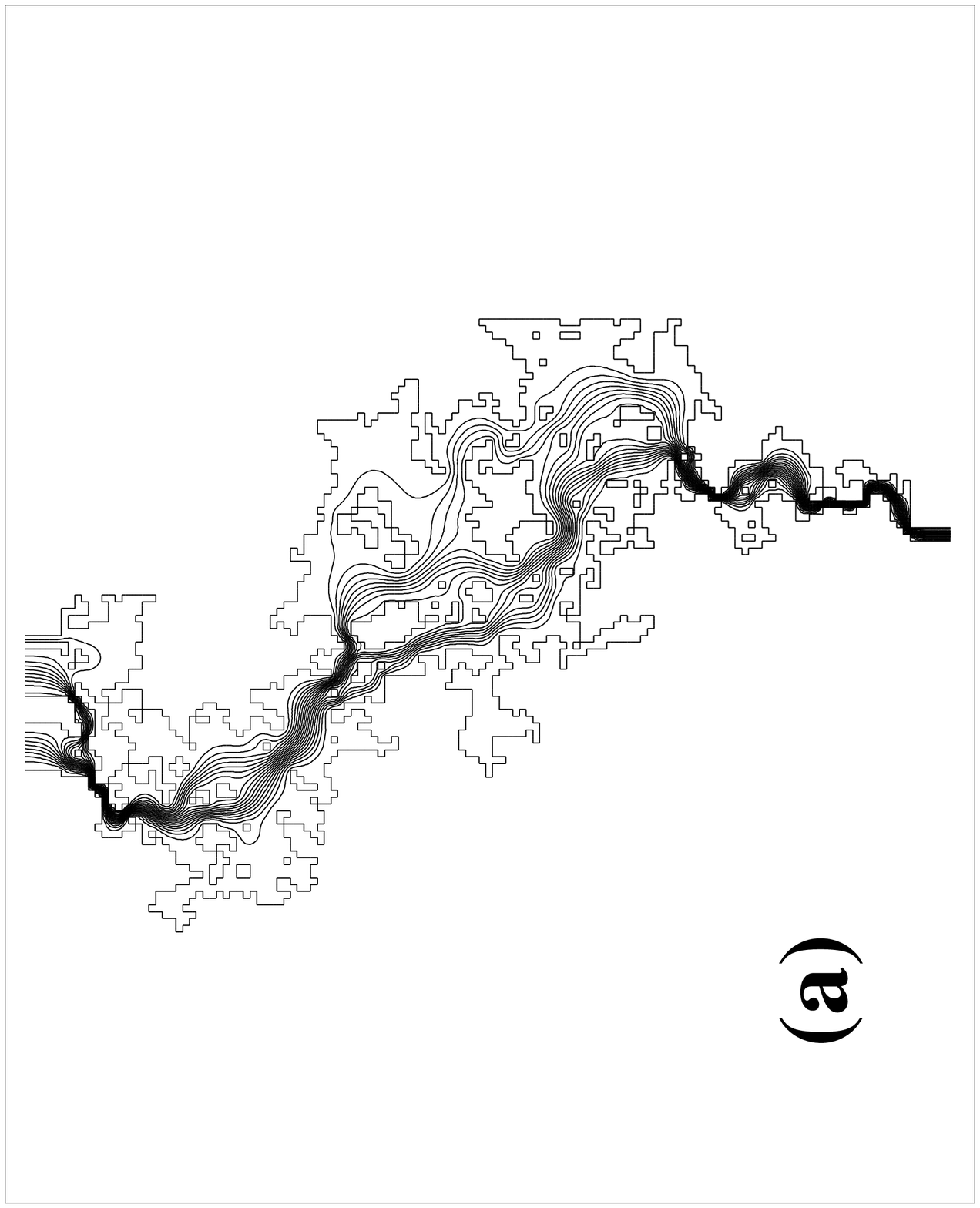}} 
   }
%\vspace*{.5cm}
\centerline{
\epsfxsize=6.cm \epsfbox{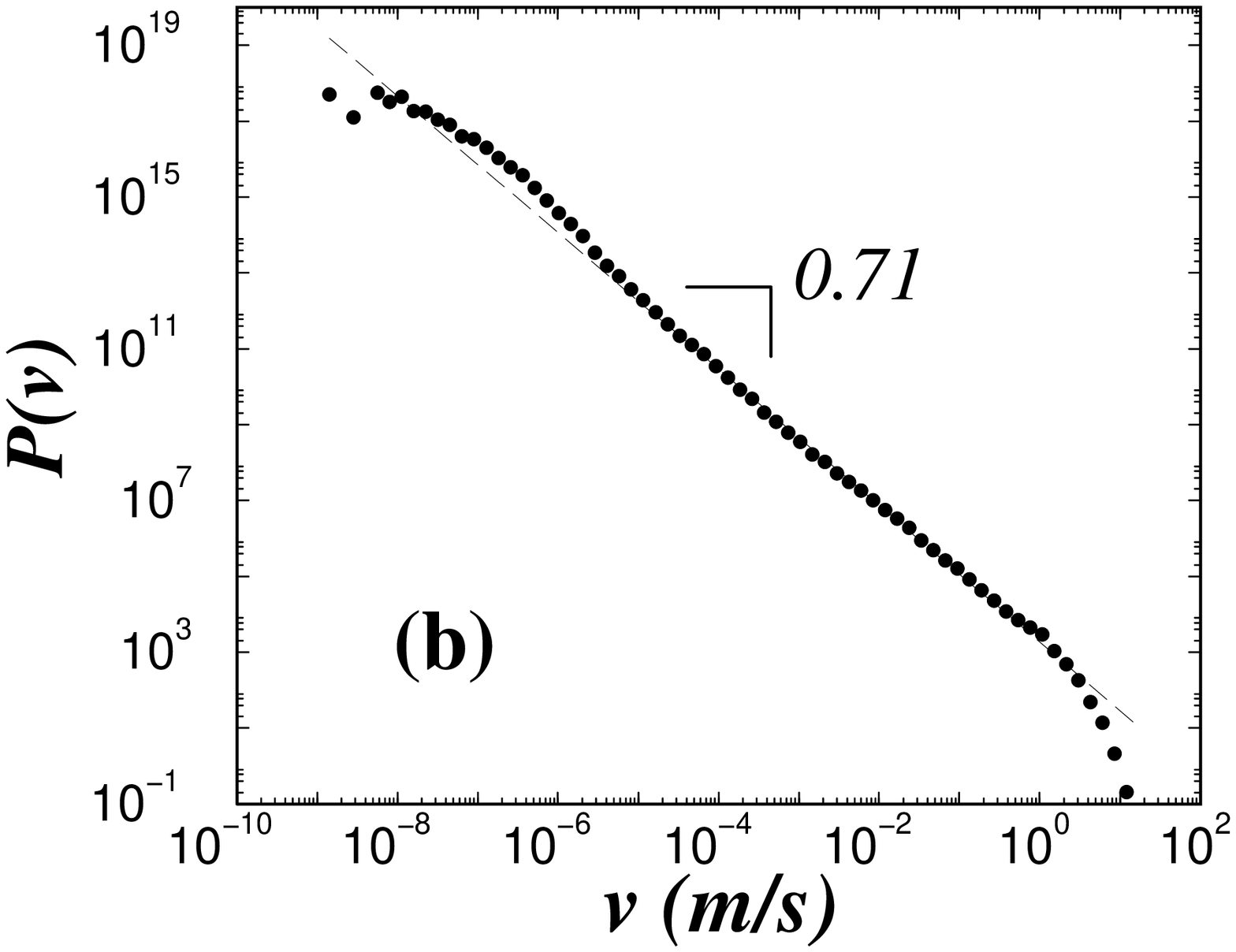} 
}
%\vspace*{.5cm}
\centerline{
\epsfxsize=6.cm  \epsfbox{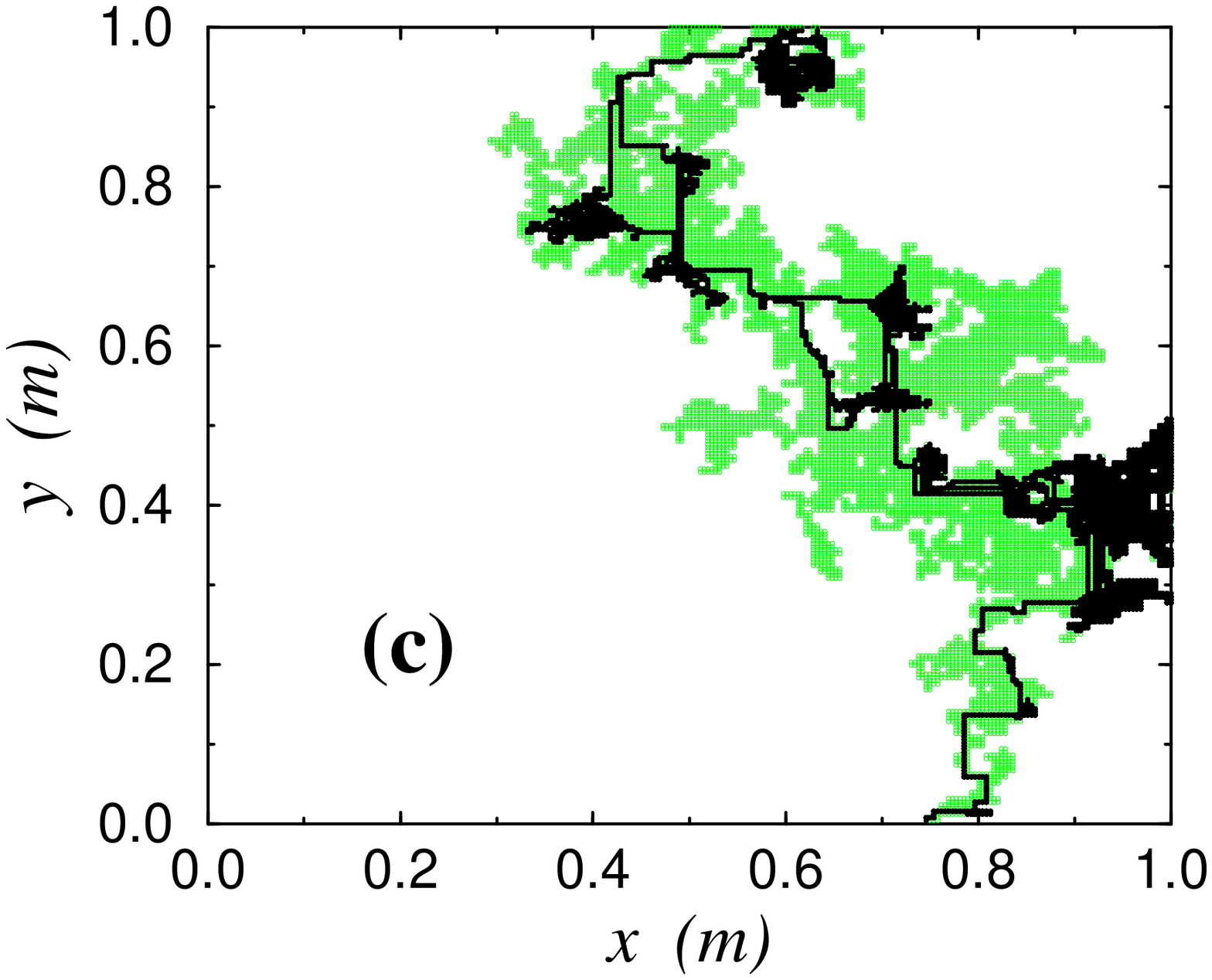} 
}
%\vspace*{.5cm}
\caption{(a) Typical stream-lines of the 
velocity field in a correlated percolation cluster.  (b) Velocity
magnitudes probability distribution averaged over 5 realizations of the
percolation clusters.  (c) Tracer diffusion in the porous medium shown
in (a), for ${\rm Pe}=1.7$.  We release five walkers and the black dots
indicate the sites visited at least one time by the walkers.}
\label{field}
\end{figure}
As a consequence, inertial effects might be locally relevant.  To avoid
this problem, we use the steady-state velocity field obtained by solving
the full set of Navier-Stokes in the percolation geometry. Then we study
the transport properties of a dynamically-neutral tracer moving in the
flow field.

We treat the competition between the effects of convection and
diffusion.  The velocity field presents a broad scale-invariant
power-law distribution of magnitude values, and we find that there are
regions of very small velocity in which the tracer can be trapped.  If
convection is important, the tracer follows the stream-lines of the
fluid. When a very small velocity region is reached, molecular diffusion
effects cannot be neglected, since by diffusion the tracer may access
the stagnant zones---where it then spends a long time. We shall see that
due to the existence of these stagnant zones, the statistical properties
of the tracer--- e.g., the first-passage time and the root mean square
displacement--- can be understood using a L\'evy walk model for the
tracer motion. The existence of L\'evy statistics is also related to the
geometrical properties of the medium---whether it is correlated or
uncorrelated in the occupancy variables of the percolation cluster.

We start by describing the disordered medium and the velocity field.
Our basic model of a porous medium is a percolation model
\cite{shlomo} modified to introduce correlations among the
occupancy units \cite{sona}. We assume the existence of correlations
because we obtain a better mathematical representation of transport
properties---such as sandstone permeability---by assuming the presence
of long-range correlations in the permeability fluctuations of the
porous rock
\cite{makse1}. The permeability of rocks such as sandstone can
fluctuate over short distances, and these fluctuations significantly
affect any fluid flow through the rock. Previous models assumed that
these fluctuations were random and without short-range correlations.
However, permeability is not the result of a simple random process.
Geologic processes, such as sand deposition by moving water or wind,
impose their own kind of correlations.

The mathematical approach we apply to describe this situation is
correlated percolation. In the limit where correlations are so small as
to be negligible \cite{shlomo}, a site at position $\vec r$ is occupied
if the occupancy variable $u(\vec r)$ is smaller than the occupation
probability $0\le p \le 1$; the variables $u (\vec r)$ are uncorrelated
random numbers with uniform distribution in the interval $[0,1]$.  To
introduce long-range power-law correlations among the variables, we
convolute the uncorrelated variables $u(\vec r)$ with a suitable power
law kernel \cite{makse2}, and define a new set of occupancy variables
$\eta(\vec r)$ with long-range power-law correlations that decay as
$r^{-\gamma}$, where $r\equiv|\vec r|$ (in the following we will set
$\gamma=0.4$).
% \cite{clusters}.

We solve the full set of Navier Stokes and continuity equations at the
percolation threshold of a square lattice with $64 \times 64$ cells and
cell edge $L=1$ m.  Grid element lengths with 1/4 of the solid cell
edge, $\ell_\parallel=\ell_\perp=\ell=L/256$, have been adopted to
discretize the governing balance equations within the pore space domain
\cite{soares}.  Figure~\ref{field}a shows a typical velocity field,
while Fig.~\ref{field}b shows the probability distribution of the
velocity magnitudes averaged over five realizations of the percolation
clusters.  We find that the data are well fit by a broad power-law of
the type \cite{soares}
\begin{equation}
P(v) \sim v^{-0.71}.
\end{equation}

Next we analyze the transport properties of a neutral tracer moving in
the fluid. We use a discrete random walk model for the tracer motion.
Following the Saffman theory of dispersion in porous media, we define
the walker motion as a competition between flow-driven convection and
molecular diffusion. To allow for comparison among different regimes of
tracer dispersion, we define a macroscopic P\'eclet number as $Pe\equiv
v_{in}\ell/D_m$, where $v_{in}=1$ m/s is the fluid velocity at the inlet
boundary of the lattice. At a given position $\vec r$ in the pore space,
we define the time scale for convection $t_c\equiv\ell/v(\vec r)$.  We
choose a convective or diffusive move, and a corresponding time step
$\Delta t$ according to:
\begin{eqnarray}
~~~t_c < t_d, \qquad{\mbox{convection}, \Delta t = t_c}\\ ~~~t_c > t_d,
\qquad{\mbox{ diffusion}, \Delta t = t_d.}
\end{eqnarray}
Here $t_d\equiv \ell^2/2D_m= {\rm Pe} \ell/(2 v_{in})$ is the
characteristic time above which diffusion effects become relevant. If
the convection move is accepted, then the tracer moves to the
nearest-neighbor site in the direction given by the velocity of the
fluid and the clock is updated according to $t \to t+t_c$.  If the
diffusion move is accepted, then the tracer moves to one of the four
nearest-neighbor positions with equal probability and the clock is
updated according to $t \to t+t_d$.

We next discuss the case of large P\'eclet number, ${\rm Pe}=1.7$, so
the value of $t_d$ is such that diffusion only occurs in regions of
small fluid velocity. Typical tracer trajectories are shown in Fig.
\ref{field}c. We see that the tracer particles perform walks with very
long straight trajectories followed by periods where they get trapped in
small velocity zones.  These ``stagnant zones'' in the pore space differ
significantly from the dangling ends of the analogous electrical problem
(i.e., the parts of the infinite cluster connected by only one site to
the backbone).  The tracer enters these regions by diffusion, and
requires a long time to escape. After escaping, the particle performs
another almost ballistic trajectory until it penetrates into the next
small velocity region. The tracer trajectory resembles a
quasi-one-dimensional channel of ``tubes and blobs.''  The ``tubes and
blobs'' picture is the analog for this problem of the traditional
``links and blobs'' picture associated with anomalous diffusion in
percolation clusters \cite{stanley,shlomo2}.

\begin{figure}
\centerline{
\epsfxsize=6.2cm \epsfbox{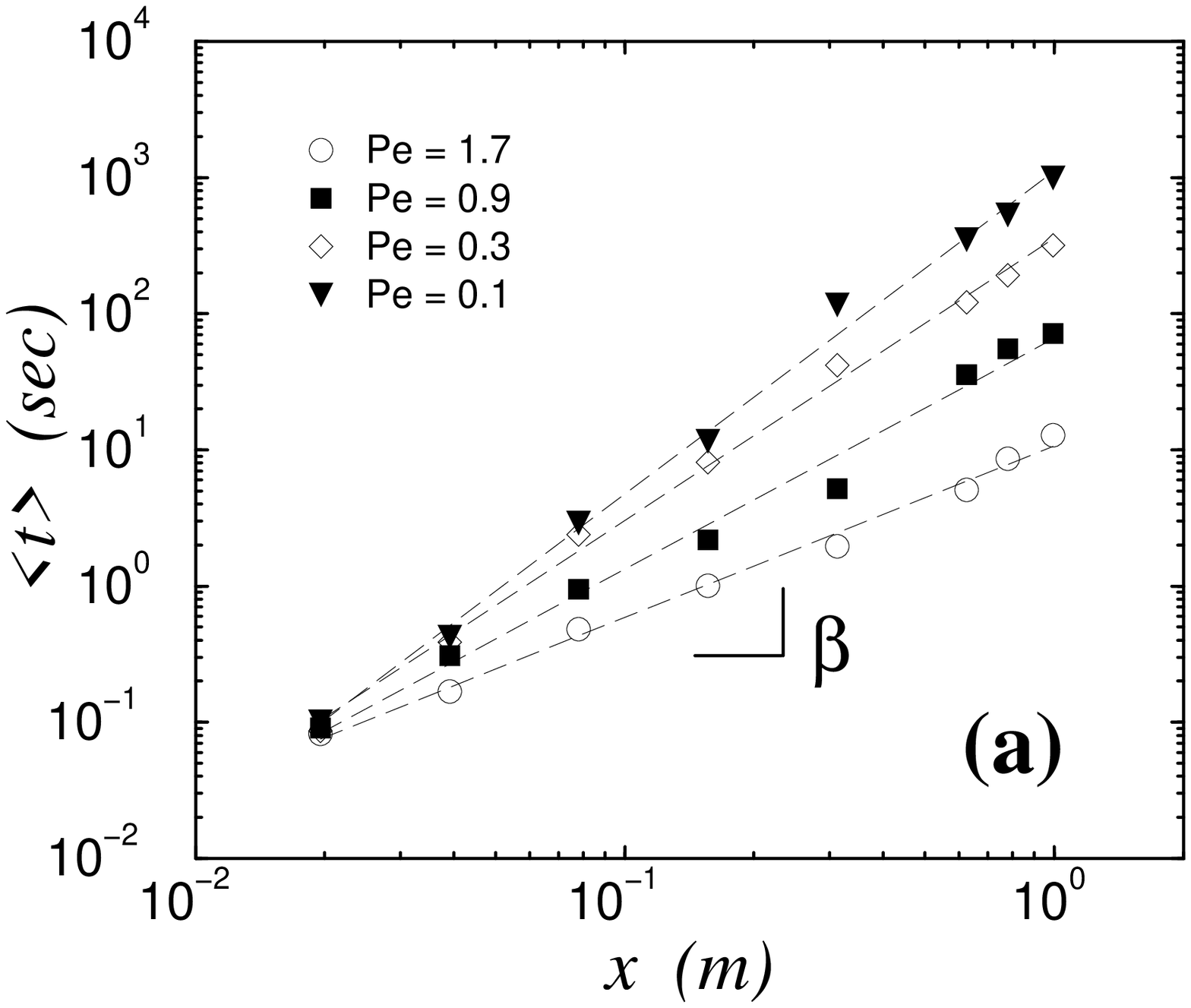} 
   }
%\vspace*{.5cm}
\centerline{
\epsfxsize=6.2cm \epsfbox{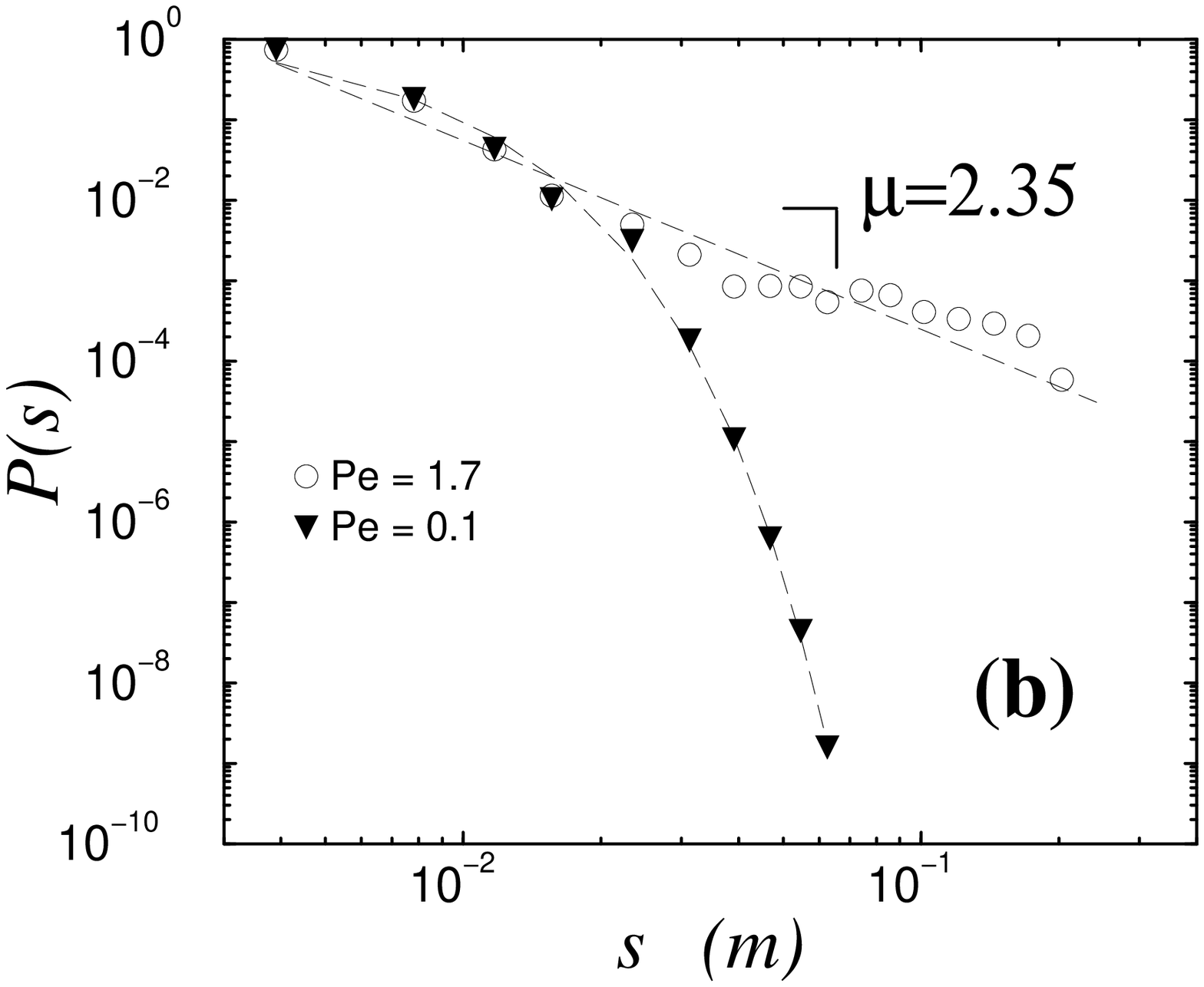} 
}
%\vspace*{.5cm}
\centerline{
\epsfxsize=6.2cm 
 \epsfbox{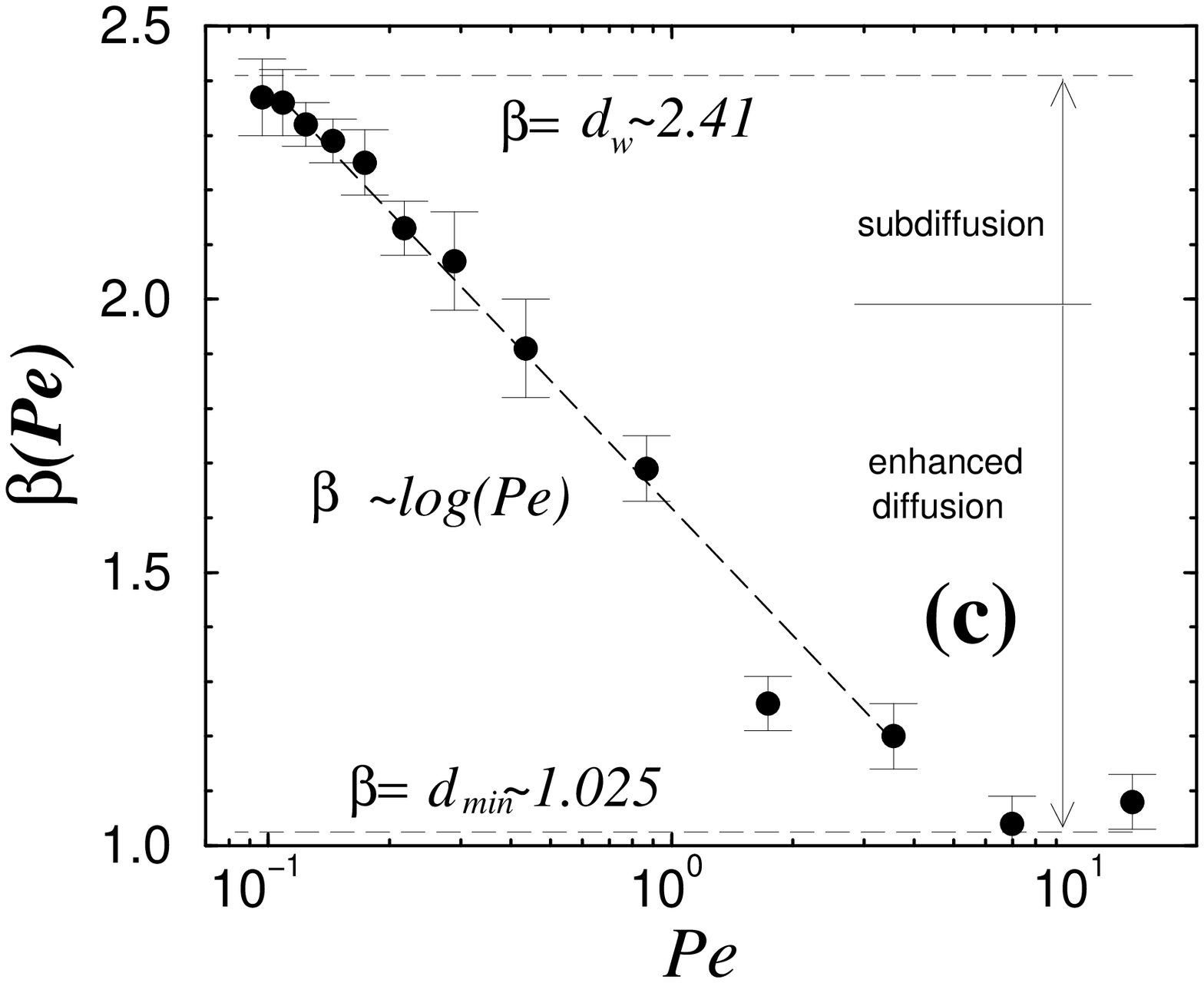} }
%\vspace*{.5cm}
\caption{(a) Transit times for different P\'eclet numbers,
averaged over five realizations of the percolation clusters.  (b)
Probability distributions of steps for the case of ${\rm Pe}=1.7$
(power-law distribution) and ${\rm Pe}=0.1$ (Gaussian distribution). (c)
Transit time exponent as a function of the P\'eclet number.
%The upper limit corresponds to the value of the
%exponent of the anomalous diffusion in correlated percolation clusters
%$d_w$. The lower limit corresponds to the minimum path exponent
%$d_{min}$, and in between those limits we find that our data can be
%approximated by a logarithmic dependence with $D_m$.
}
\label{levy}
\end{figure}

We analyze the transit time, i.e., the average time required for the
tracer to traverse a given distance $x$ from the inlet line, $0<x<L$,
for different P\'eclet numbers. We find (Fig.~\ref{levy}a) that the
transit times follow a power law
\begin{equation}
\langle t\rangle\sim x^\beta
\end{equation}
where $\beta\simeq1.26$ when ${\rm Pe}=1.7$.

In the ``tubes and blobs'' picture, we define a tube as the set of steps
taken by the tracer following a fixed direction, and we analyze the
statistical distribution of the tube length $s$.  In stagnant zones
where diffusion is dominant, the tracer is expected to change direction
every time step, so that $s\simeq \ell$.  In regions in which convection
dominates, the tracer moves in ballistic trajectories limited only by
impermeable obstacles.  Since long-range correlated clusters are very
compact (Fig. \ref{field}a), we expect $s\gg
\ell$ and the tube length distribution to be broad.  Both
expectations are corroborated by our calculations.

Figure~\ref{levy}b shows the probability distribution $P(s)$ for two
different values of ${\rm Pe}$.  We find that for sufficiently large
${\rm Pe}$, the step lengths follow a scale invariant power law
distribution
\begin{equation}
P(s) \sim s^{-2.35} \qquad \mbox{[${\rm Pe}=1.7$]},
\label{pl}
\end{equation}
while for small ${\rm Pe}$ values, when diffusion is the dominant
mechanism in the entire pore space, the distribution is Gaussian
\begin{equation}
P(s) \approx e^{-\frac{1}{2}(s/s_0)^2} \qquad \mbox{[${\rm Pe}=0.1$]},
\label{gauss}
\end{equation}
with $s_0$ a characteristic jump length.

In case (\ref{pl}), the step lengths statistics can be considered as a
L\'evy walk, i.e., a random walk process in which the jump distribution
is a power law
\cite{shlesinger1}
%,shlesinger3,weiss-bouchaud}

\begin{equation}
P(s) \sim s^{-\mu}.
\label{levy-stat}
\end{equation}

A random walker with a distribution (\ref{levy-stat}) travels a typical
distance $r\sim t^{2-\mu/2}$, when $2<\mu<3$. Thus, the transit time for
a L\'evy walker with jump statistics given by (\ref{levy-stat}) is
\cite{shlesinger1}
\begin{equation}
\langle t\rangle \sim x^{2/(4-\mu)}.
\label{levy-t}
\end{equation} 
For $\mu=2.35$--- the value we find in our simulations for ${\rm
Pe}=1.7$--- we obtain $\langle t\rangle \sim x^{1.21}$, which agrees
with the scaling found when we calculate the transit time directly,
$\langle t\rangle \sim x^{1.26}$ from Fig.~\ref{levy}a, and confirms the
validity of the L\'evy walk picture as an accurate description of the
tracer motion at large ${\rm Pe}$.

The transit time exponent $\beta$ is not universal and depends on ${\rm
Pe}$ (Fig. \ref{levy}c).  In fact we find that the L\'evy statistics
approximates well the value of $\beta$ in the entire enhanced diffusion
regime $1<\beta<2$, while in the sub-diffusion regime $\beta>2$, the
L\'evy statistics Eq. (\ref{levy-t}) ceases to be valid.  Moreover, we
expect two limiting regimes. If convection dominates completely
(mechanical dispersion), then the tracer should follow the minimum path
along the spanning percolation cluster. The minimum path length
$\ell_{\mbox{\scriptsize min}}$ scales as $\ell_{\mbox{\scriptsize
min}}\sim x^{d_{\mbox{\scriptsize min}}}$ where $d_{\mbox{\scriptsize
min}}$ is the fractal dimension of the minimum path distance between two
points separated by a linear distance $x$
\cite{shlomo}.  If the tracer  moves with a constant
velocity, we can identify the minimum path distance with the transit
time, so $\beta=d_{\mbox{\scriptsize min}}$.  This is the lower limit of
the transit time exponent, and we confirm this prediction since we
obtain $\beta \stackrel{>}{\sim} d_{\mbox{\scriptsize min}}$ when ${\rm
Pe}$ is large (Fig.~\ref{levy}c) \cite{dmin}.

The other limit at larger diffusivities--- the anomalous diffusion case
\cite{shlomo2}--- corresponds to the regime dominated completely by
diffusion, and the transit time scales as
%\begin{equation}
$\langle t\rangle \sim x^{d_w},$
%\end{equation}
 where $d_w$ is the random walk fractal dimension. The value $d_w$
depends on the degree of correlation, with $d_w=2.87$ for the
uncorrelated percolation limit \cite{shlomo} and $d_w=2.41$ \cite{sona}
for the correlated percolation problem we study ($\gamma = 0.4$). We see
that the limiting cases of our calculations agree with these predictions
(Fig.~\ref{levy}c). Between these two limiting cases, we find that the
transit time exponent can be approximated by
% a logarithmic dependence on
%the P\'eclet number
\begin{equation}
\beta({\rm Pe})\sim \log({\rm Pe}).
\end{equation}

We also perform simulations on uncorrelated percolation clusters. We
find a enhanced diffusion regime and a sub-diffusion regime as well.
%
%, with the two limiting cases corresponding to the minimum path at large
%${\rm Pe}$ and to the anomalous diffusion case at small ${\rm Pe}$ values.  
However, due to the tortuosity of the uncorrelated percolation clusters
at the threshold, the distribution of steps is not a scale-free
power-law, as we find in the case of enhanced diffusion in correlated
clusters Eq.  (\ref{pl}).  Thus, we conclude that the L\'evy statistics
found in the case of dispersion in correlated clusters is a by-product
of the dynamical properties of the tracer moving in a broadly
distributed velocity field plus the geometrical properties of the
particular porous medium treated here. The compact features of
long-range correlated percolation clusters allows the tracer to perform
large ballistic steps without encountering obstacles during the random
walk process.

In summary,
%with detailed descriptions of the porous geometry and 
%fluid flow, we show that it is possible to capture certain hydrodynamic 
%features of the dispersion phenomenon that have not been studied before, 
%at least from a microdynamical perspective. Indeed, 
we find that, at sufficiently large P\'eclet numbers, there is a regime
of dispersion for correlated porous media in which the trajectory of the
tracer particle should be better described by a L\'evy statistics Eq.
(\ref{pl}) instead of the Gaussian behavior Eq. (\ref{gauss}).
Interestingly, this fact should be relevant to elucidate the mass and
momentum transport mechanisms responsible for the dispersion regime
called ``holdup dispersion'' \cite{review}. Tracer experiments indicate
that this regime of strong dependence between dispersion measurements
and P\'eclet number is typical of percolation-like porous materials.

 \end{multicols} 
\begin{references}
\bibitem{dmin}
Notice that the value $d_{\mbox{\scriptsize min}} \simeq 1.025$
corresponds to the correlated percolation value
\protect\cite{sona}, 10\% smaller than the value $d_{\mbox{\scriptsize
min}}\simeq1.135$ for the uncorrelated percolation problem
\protect\cite{shlomo}.
\end{references}
\end{document}